\documentclass[prb,superscriptaddress]{revtex4}
\usepackage{graphicx}
\usepackage{dcolumn}
\usepackage{bm}
\usepackage{amsmath}
\usepackage{ulem}
\def\ffi{\varphi}
\def\bL{{\bf L}}
\def\br{{\bf r}}
\def\bk{{\bf k}}

\def\bq{{\bf q}}
\def\be{{\bf e}}
\def\vkk{\frac{1}{|k-k'|^{D-1}}}
\def\Vkk{\frac{1}{|k-k'|}}
\DeclareMathOperator{\asinh}{asinh}
\DeclareMathOperator{\Span}{Span}

\def\simge{\mathrel{ 
       \rlap{\raise 0.511ex \hbox{$>$}}{\lower 0.511ex \hbox{$\sim$}}}}
\def\simle{\mathrel{
       \rlap{\raise 0.511ex \hbox{$<$}}{\lower 0.511ex \hbox{$\sim$}}}}
\begin{document}

\title{Metal-insulator transition in the Hartree-Fock phase diagram of the
fully polarized homogeneous electron gas in two dimensions} 
\author{B. Bernu}
\affiliation{LPTMC, UMR 7600 of CNRS, Universit\'e P. et M. Curie, Paris, France}
\author{F. Delyon}
\affiliation{CPHT, UMR 7644 of CNRS, \'Ecole Polytechnique, Palaiseau, France}
\author{M. Duneau}
\affiliation{CPHT, UMR 7644 of CNRS, \'Ecole Polytechnique, Palaiseau, France}
\author{M. Holzmann}
\affiliation{LPTMC, UMR 7600 of CNRS, Universit\'e P. et M. Curie, Paris, France}
\affiliation{LPMMC, UMR 5493 of CNRS, Universit\'e J. Fourier, Grenoble, France}

\date{\today}
\begin{abstract}
We determine numerically the ground state of the two-dimensional, fully polarized electron gas within the Hartree-Fock approximation without imposing any particular symmetries on the solutions.
At low electronic densities, the Wigner crystal solution is stable, but for higher densities ($r_s$ less than $\sim  2.7$)  we obtain a ground state of different symmetry: the charge density forms a triangular lattice with about $11\%$ more sites than electrons.  
We prove analytically that this conducting state with broken translational symmetry has lower energy than the uniform Fermi gas state in the high density region giving rise to a metal to insulator transition.  
\end{abstract}
\maketitle
\section{Introduction}

The two-dimensional homogeneous electron gas is one of the fundamental models in condensed matter physics. 
Despite its simplicity -- the system consists of electrons interacting through a $1/r$-potential to which a uniform positive background is added for charge neutrality  -- the phase diagram at zero temperature  is nontrivial\cite{Tanatar,Bernu,Waintal}. 
In general, it is given in terms of the dimensionless parameter $r_s=1/\sqrt{\pi na_B^2}$, where $n$ is the electronic density and $a_B=\hbar^2/(me^2)$  the Bohr radius (see Section \ref{notations} for notations and units). 
At low density (large $r_s$), the potential energy dominates over the kinetic energy and the system
forms a triangular lattice, the Wigner crystal (WC), whereas in the high density region 
($r_s \to 0$) the kinetic energy favors a uniform Fermi gas (FG) phase\cite{Tanatar} (the simplest state given as the determinant of the plane waves with wave vector $\bk$ of modulus smaller than $k_F$).
The energy of the FG is known analytically.
Already Wigner\cite{Wigner}  argued that the unpolarized FG is unstable even in the limit $r_s \to 0$.
Later, Overhauser claimed the instability of the unpolarized WC with respect to spin-density waves, even within the Hartree-Fock approximation (HF)\cite{Overhauser}.
It has further been  conjectured that the Coulomb potential prevents any  first order transition between the WC and a FG\cite{Spivac}. 
Despite these rather general instability theorems, there are few quantitative calculations of the true ground state of the electron gas within HF\cite{Giulani}.
 A previous HF study \cite{Needs} of the two and three dimensional electron gas compares the FG energy with the energy of various states with imposed crystal symmetries. 
For the polarized two-dimensional gas, they find lower energies for a crystal for $r_s$ larger than $2$.
Only recently an unrestricted  HF study of the unpolarized three-dimensional electron gas was performed which proposes a more complicated structure of a ground state with spin-density waves in the high density region\cite{Shiwei}.

Indeed, establishing  the precise HF phase diagram of the electron gas influences the correlation energy  estimations, since by definition the many-body correlation effects must be  evaluated with respect to 
the true HF ground state. Further, even in more advanced technics, the antisymmetry of the wavefunctions is in general provided by  a single Slater determinant.

In the present study, we consider the fully polarized two-dimensional electron gas, analytically, and numerically.
Section \ref{numerics} summarized our numerical results.
At low densities, $r_s \gtrsim 2.7$, our simulations always lead to a WC. 
For higher densities and large enough number of electrons, $N$, the solution is neither a FG nor a WC: the density modulation corresponds to a partially occupied crystal of different symmetry compared to the WC phase.
As the number of sites is larger than $N$, we refer to this solution as a metallic phase.
Details of our numerical methods are given in section \ref{algo}.

In Section \ref{notations} we remind some definitions and notations particularly
used in the following.
Section \ref{bounds} is devoted to derive rigorous, analytical upper bounds on the energy of the metallic phase. These bounds are obtained in the limit $r_s \to 0$ where the calculation is simplified by the long range behavior of the interaction potential. 
In the conclusion, Section \ref{conclusion}, we briefly summarize the results of the paper and discuss
their relevance. 

The new HF solutions discussed in this paper within the HF approximation open a new perspective for the qualitative understanding of the experimental observed  metal to insulator transition\cite{Kravchenko94} and should be considered in studies beyond the HF approximation.
The possibility that the experimental findings are driven by interaction effects  -- and not  by disorder -- was recently adressed in Refs~\cite{Pankov,Kotliar}
considering an extended Hubbard model.

\section{Numerical results}
\label{numerics}
The N-body Hamiltonian, $H=K+V$, contains the kinetic energy $K$ and the $1/r$-periodic Coulomb potential $V$ where a uniform positive charge background is subtracted. 
Within the HF approximation, the search of the true ground state of the quantum many-body system is reduced to the simpler problem of finding the lowest energy states in the subset of the Slater determinants (see Eq. \ref{energy}). 
Let $\Phi=\psi_1\wedge\cdots \wedge\psi_N$ be the Slater determinant associated with the single particle states $\{\psi_i\}$ and $E(\psi_1,\psi_2,\dots,\psi_N)$ the corresponding energy expectation value.

By a kind of descent method described in the following section, we numerically study systems with up to $500$ electrons at densities  corresponding to $r_s=1$ up to $r_s=30$.
Since we expect the electrons to crystallize on a triangular lattice at  low densities, we choose periodic conditions compatible with this geometry.
Thus, the unit cell  of the periodized system is given in terms of  two vectors $\left\{\bL_1,\bL_2\right\}$ of length $L$ and with an angle of 60 degrees between both; the  volume of the unit cell is $\Omega= L^2\sqrt{3}/2$. 
We have restricted our study to system sizes which are compatible  both with the triangular lattice and with a closed shell occupation in $k$-space.
Any triangular crystal with unit-cell vectors $\left\{\be_1,\be_2\right\}$ is compatible with the boundary conditions if it satisfies $\bL_1=l \be_1+m\be_2$ and $\bL_2=-m \be_1+(l+m)\be_2$, where $(l,m)$ are two non-negative integers. 
The number of sites of the lattice is given by $N_c=\det(\bL_1,\bL_2)/\det(\be_1,\be_2)=l^2+m^2+lm$.
\begin{figure}
\begin{center}
\includegraphics[width=0.45\textwidth]{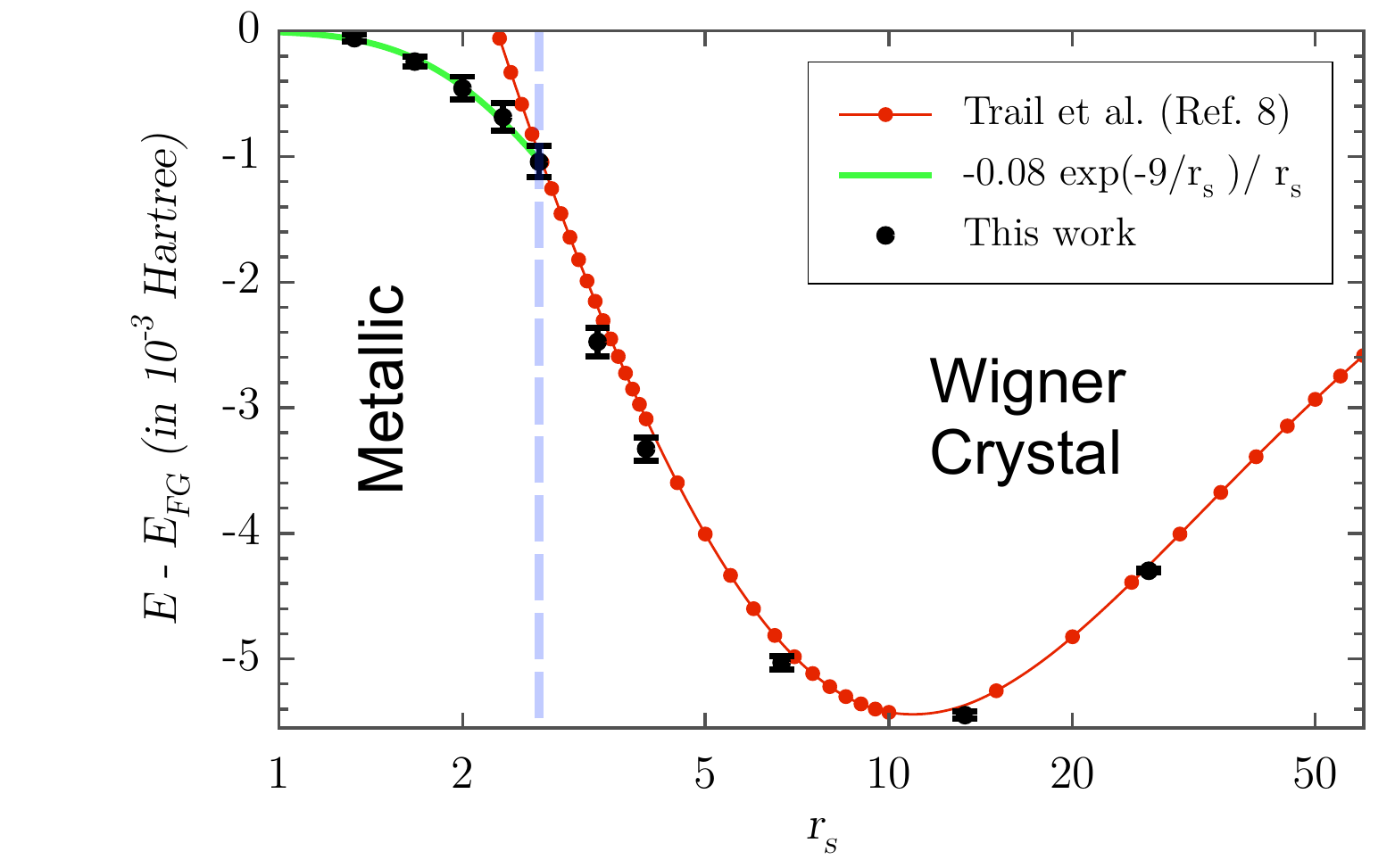}
\caption{Extrapolated energies $E_{\infty} - E_{FG}$ (milli-Hartree units) versus $r_s$. Points with error bars:  present calculations, full line (red): data of Ref.~\cite{Needs}, full line (green): fit to present results for ($r_s\lesssim 2.7$), vertical dash line (blue): $r_s\sim2.7$.
}
\label{FIG-Comparison}
\end{center}
\end{figure}

In Figure \ref{FIG-Comparison}, we report the energies of the obtained HF ground state $E_{\infty}(r_s)$, extrapolated to the thermodynamic limit, as a function of $r_s$. 
In the low density region, for $r_s>3$, we obtain good agreement with the results of Trail {\it et al.}\cite{Needs}, which imposed a ground state build  as a complete band of Bloch wave functions of the triangular WC lattice. 
But for smaller $r_s$ we find  lower energies which remain also below the FG energy down to $r_s=1$.

\begin{figure}
\begin{center}
\includegraphics[width=.7\textwidth]{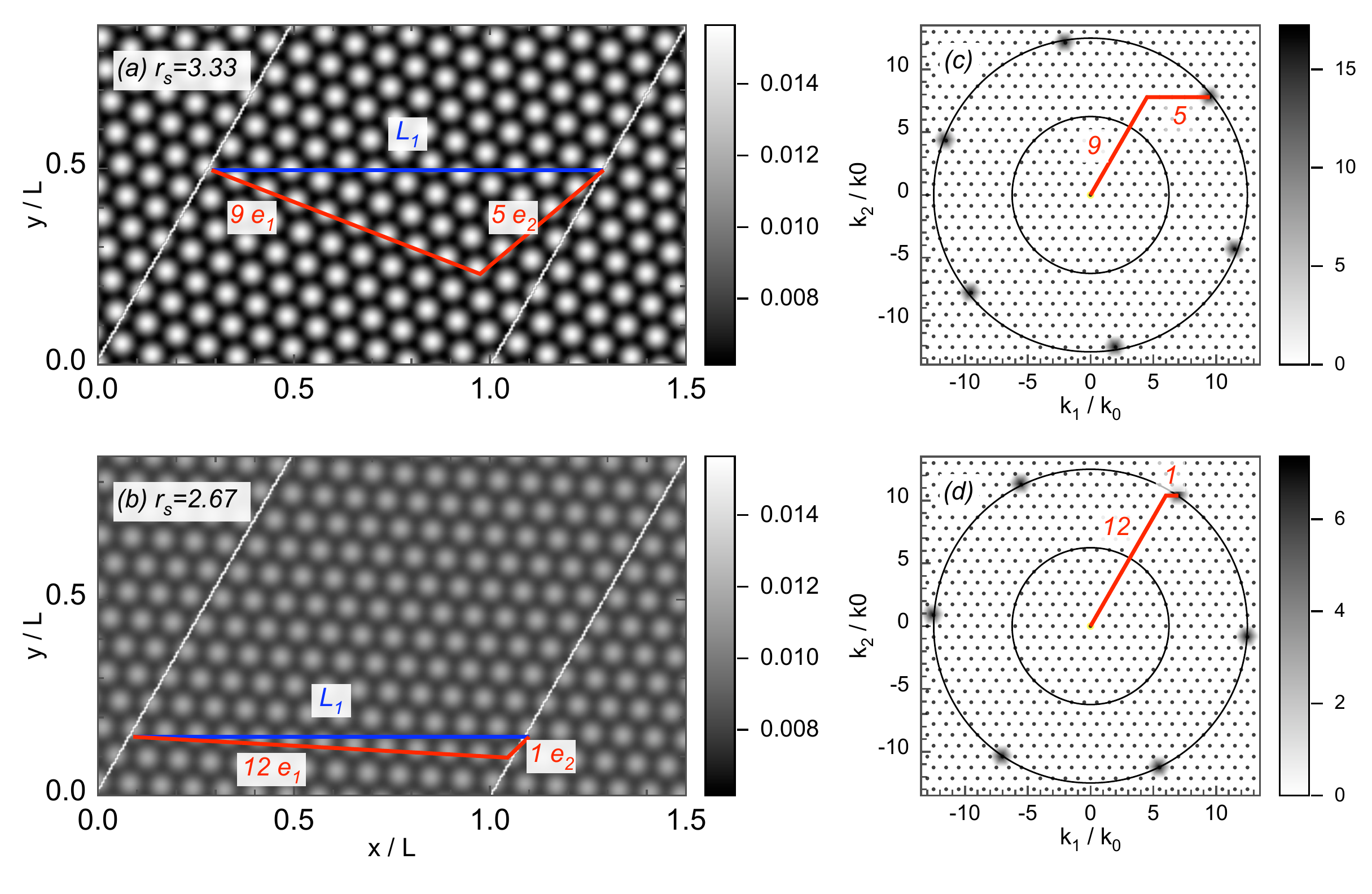}
\caption{Left : charge density $\rho(\br)/\langle\rho\rangle-1$, with $\rho(\br)=\sum_{i=1}^N |\ffi_i(\br)|^2$. 
Right: $\tilde\rho(\bk)$, the Fourier transform of the charge density, where $\tilde\rho(0)= N$ has been removed. The grid-points are thoose compatible with the periodic conditions.
Top: the number of maxima is $N_c=9^2+5^2+9\times5=151$. 
Bottom:  the number of maxima is $N_c=12^2+1^2+12\times1=157$. 
Gray levels corresponds to the same density in both figures.
Colored lines correspond to $\bL_1=l \be_1+m\be_2$, where the numbers stands for $l$, $m$ (see text).}
\label{FIG-chargedensities}
\end{center}
\end{figure}

The Figure \ref{FIG-chargedensities} (a) shows the typical charge density for $r_s\approx 3$.
In this case we have a triangular lattice with exactly  $N$ sites. 
The Figure \ref{FIG-chargedensities} (c) shows the Fourier transform of the charge density. 
The support of the Fourier transform is the six-fold star corresponding to the triangular lattice of the charge density with $N_c=N$.

For $1<r_s<2.7$, we find new kind of ground states (see Figure \ref{FIG-chargedensities} (b)): the support of the Fourier transform ((see Figure \ref{FIG-chargedensities} (d)) is still a six-fold star corresponding to a triangular lattice, but this triangular lattice has a number of sites $N_c$ larger than  $N$. 
The system lowers its energy by delocalizing the electrons on a denser lattice with more sites than electrons (Figure \ref{FIG-chargedensities}). 
This denser lattice is characterized by integral numbers $(l',m')$ different from the WC lattice $(l,m)$. 
For some system sizes $N$, the maxima of the Fourier transform correspond to various couples of $\{l',m'\}$ leading to different number of lattice sites, $N_c=l'^2+m'^2+l'm'$.  
In any case, the system looks like a periodic crystal with an incomplete band in contrast to the WC solutions of fully occupied bands, studied in Ref. \cite{Needs}. 
We refer to this solution as the metallic phase.  
However, as $r_s$ approaches zero, the energy gain of this metallic crystal compared to the FG gets more and more tiny. 
At the same time, $N_c$ is either constant or increases when $r_s$ decreases (apart for a few exceptions).
At $r_s<1$, the FG solution is stable for our finite system sizes ($N \le 500$). 

Now we would like to understand the nature of the Slater determinants in the metallic phase.
A Slater determinant is obtained as a set of $N$ orthonormal single particle wavefunctions $\psi_i$.
Only the space generated by the $\psi_i$'s is relevant, and in order to understand the numerical results we need to choose a canonical representation of the $\psi_i$'s.

Let $\{\phi_i\}_{i=1\ldots N}$ be a basis corresponding to some indexation of the plane waves associated to the wave vectors $\bk_i$ of the Fermi sphere.
As $r_s$ is small, the space generated by the $\psi$'s becomes close to the space generated by the $\phi$'s.
Let $M$ be the square matrix defined by $M_{ij}=\left<\phi_i| \psi_j\right>$ which measures the overlap of the two Slater determinants. 
The Singular Value Decomposition (SVD)  of $M$ is $M=U\sigma V$ where $U$ and $V$ are unitary matrices and $\sigma$ is a diagonal positive matrix. 
Then the orthonormal set  $\{\psi'_i\}_{i=1\ldots N}$ defined by:
\begin{eqnarray}
\psi'_i=\sum_k \overline{UV}_{ik}\psi_k
\end{eqnarray}
is a basis of $\Span(\{\psi_i\})$ close to the basis $\{\phi_i\}_{i=1\ldots N}$ of $\Span(\{\phi_i\})$ in the sense that
$\left<\phi_i| \psi'_j\right>=(\overline U\sigma U^T)_{ij}\approx \delta_{ij}$ as soon as $\sigma_i$'s is close to one.

From now on, we assume that the single particle wavefunctions, $\psi_i$, have been chosen in this way.

\begin{figure}
\begin{center}
\includegraphics[width=0.4\textwidth]{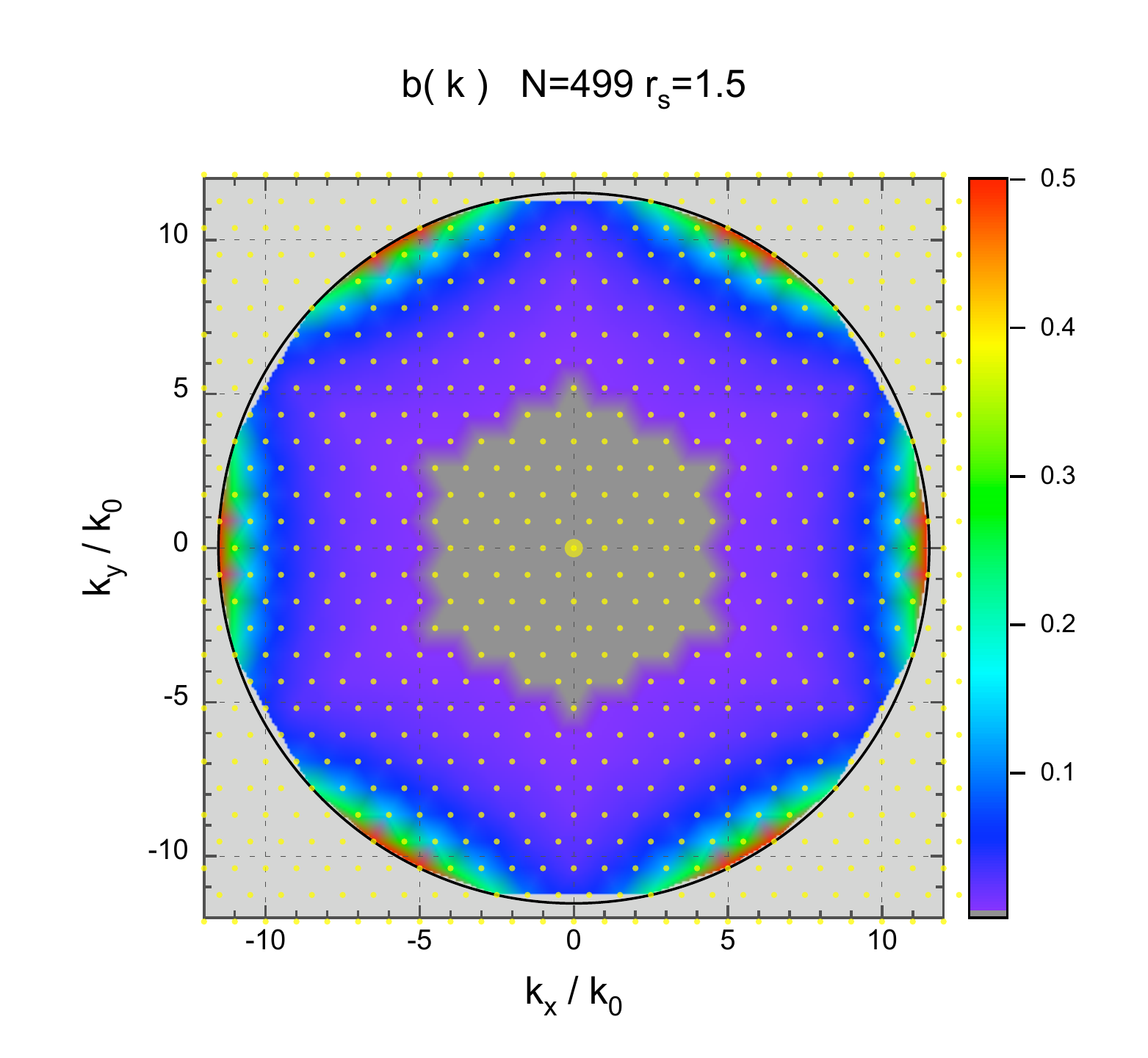}
\caption{Numerical values of $b_k$ for 499 electrons in two dimensions at $r_s=2$.
}
\label{b499}
\end{center}
\end{figure}

Thus if $r_s$ is not too large, $\psi_i$ is close to $\phi_i$ (at least for $i$ associated to a wave vector not too close to the Fermi surface) that is $\psi_i(\bk_i)$ is close to one. 
Thus, the largest amplitude of $\psi_i(\bk)$ is for $\bk=\bk_i$, and Fig.\ref{b499} represents the next largest amplitude of $\psi_i$, that we denote $b_{\bk_i}$, for 499 electrons at $r_s=2.7$ in 2D.

For $|\bk_i| \approx k_F$,  $\psi_i$ has essentially only two nonzero components: one at $\bk=\bk_i$ and the other one $b_{k_i}$
at the vector $\bk=\bk_i+\bq_i$  where $\bq_i$ is the vector of the six-fold star of Figure \ref{FIG-chargedensities} (d)  such that 
$\bk$ is close to the Fermi surface.
This condition can only be satisfied for a set of $\bk_i$ closed to a six-fold star as we see on Figure \ref{b499}.

One can understand why metallic states should exist at small $r_s$ in the thermodynamic limit.
Let us replace a plane wave state $\bk$ of the FG ($\| \bk \| \le k_F$) by a superposition of two plane waves with wavevectors $\bk$ and $\bk+\bq$ ($\| \bk +\bq\| > k_F$). 
Choosing $\bq$ on the six-fold star of a triangular lattice we certainly obtain a gain in potential energy.
The increase of kinetic energy is minimized if $\|\bk\|\sim k_F$ and $\| \bk +\bq\| \sim k_F$.
Then, the number of solutions for $\bk$ is optimal if $\|\bq\|\sim 2 k_F$.
This solution corresponds  to a triangular lattice of length $L_c=2 \pi/(\sqrt{3} k_F)$ in real space leading to a unit cell of volume $\Omega_c=\sqrt{3}L_c^2/2$. 
Since the system is contained in the volume $\Omega= \sqrt{3}L^2/2$, we will obtain $N_c=\Omega_c/\Omega$ lattice sites, or
\begin{equation}
\label{EQ-Nc}
N_c = \frac{2\sqrt{3}}{\pi} N \approx 1.1 N, 
\end{equation}
where we have used $n=N/\Omega=k_F^2/(4\pi)$.

We will elaborate this argument into an analytical proof of an upper bound on the ground state energy in section \ref{bounds}.

\section{The descent method}
\label{algo}
In this section we provide some details about the descent method used to obtain numerically
 the HF ground states of the electron gas.
The variation of the total energy ${\cal E}(\psi_1,\psi_2,\dots,\psi_N)$ with respect to a variation 
of the single particle state
$\delta\psi_i$
is  given by
\begin{eqnarray}
\delta {\cal E}&=&\sum_i \left<h_\Psi\psi_i | \delta\psi_i\right>+
\sum_i \left<\delta\psi_i | h_\Psi\psi_i\right>.
\end{eqnarray}
where $h_\Psi$, the so-called HF Hamiltonian, is a single particle operator depending on the full state $\Psi$ (not on the particular choice of the $\psi_i$'s). Extremal states must satisfy the following
equation
 \begin{align}
\label{EqH}
h_\Psi\psi_i=\sum_j C_{ij}\psi_j,
\end{align}
where $C_{ij}$ are the Lagrange coefficients associated with the normalization constraint 
$\left<\psi_i|\psi_j\right>=\delta_{ij}$. Conversely, if $\Psi=\psi_1\wedge\cdots \wedge\psi_N $ is not an extremum, we have
 \begin{align}
\label{EqH1}
h_\Psi\psi_i=\sum_j C_{ij}\psi_j+\theta_i
\end{align}
where the $\theta_i$'s satisfy $\left<\psi_i|\theta_j\right>=0$, $\forall i,j$.
Within the steepest descent method
one chooses first a $N\times N$ unitary transformation $A=(a_{ij})$ such that one obtains
\begin{align}
\left<\psi'_i| \theta'_j\right>=0, \quad 
\left<\theta'_i | \theta'_j\right> \propto \left<\psi'_i | \psi'_j\right>=\delta_{ij},
\quad  \forall i,j
\end{align}
for the transformed single particle states
$\theta'_i=\sum_ja_{ij}\theta_j$, $\psi_i'=\sum_ja_{ij}\psi_j$.
The energy ${\cal E}(\psi_1+\lambda\theta_1,...,\psi_N+\lambda\theta_N)$ can be expressed as a sum of rational fractions whose numerators and denominators are polynomials of order four, at most. Thus, it is possible to find the best $\lambda$ and to iterate the process until a stationary state is reached.

In fact, this method has the same drawbacks as the steepest decent method in linear optimization problems; in general, it converges slowly. 
For linear problems, conjugate gradient methods are preferable \cite{CG, FDMD}. 
However, since the HF states do not form  a linear space, the genuine conjugate gradient method does not apply here. We have therefore adapted a variant of this method to the non-linear case.
Let $\eta_i$ be the previous variation $\delta\psi_i$,  and $\theta_i$ is obtained by Eq.\,(\ref{EqH1}).
We then  compute ${\cal E}(\psi_1+\lambda\theta_1+\mu\eta_1,\cdots,\psi_n+\lambda\theta_N+\mu\eta_N)$ for
six values of the pairs $\{\lambda,\mu\}$ in order to  approximate ${\cal E}$ by a polynomial of order two in 
$\lambda$ and $\mu$. 
Minimizing the polynomial with respect to $\lambda$ and $\mu$, we obtain the new changes of the single particle states, $\delta\psi_i$, and the corresponding energy change. 
This process is iterated until the relative variation of the energy, $\delta {\cal E}/{\cal E}$, is sufficiently small.
 
 We compute the wavefunction on a $N_g\times N_g$ grid, the fast Fourier transform is used to switch between real and reciprocal space\cite{K-Wigner}.
We have systematically checked  the convergence of the solution  with respect to the grid size.
For the FG ground state, convergence is reached once all $k$-vectors up to $2k_F$ are represented in the grid ($N_g\sim 4\sqrt{N/\pi}$).
At larger $r_s$,  in the WC phase, the wave functions are essentially Gaussians\cite{VAR-GAUSSIAN}.
The width $\delta$ of the Gaussians scales as $\delta/L \propto (r_s N)^{-1/2}$.  
For a correct resolution of the Gaussians we need $L/N_g \propto \delta$, so that the number of grid points increases  at low densities,  $N_g \propto (N r_s)^{1/2}$.
Convergence is reached for $N_g=32$ (resp. 64, 128) for $N\le43$ (resp. $N\le200$, $N\le500$) up to $r_s=30$. 
Whenever the number of grid points is chosen too small, solutions without any particular symmetries are obtained.

We have further studied the influence of the initial state on the final solution, by choosing different types of wavefunction for  initialization: a WC state, a converged state stored at larger or lower $r_s$, a  state initialized with random numbers, or a ``metallic state'' as described above.

Typically, the energies decrease exponentially with the number of iterations. 
The decrease in energy during transitions to a different symmetry is in general much smaller than the convergence within the same symmetry.
We have often seen energy plateaus with changes of relative energy $\lesssim 10^{-4}$ just before the occurrence of a transition to a completely different state.
For system sizes up to $N=151$, the minimization is continued until a relative precision of $10^{-12}$ is reached, and for larger $N$ a relative precision of $10^{-5}$ is used. 
\section{Energy of the polarized electron gas for a Slater state}
\label{notations}
In this section we set our notations and recall the basic formulas of the electron gas.
We consider the Hamiltonian of $N$ electrons in a 2D or 3D square box of volume $\Omega$ with periodic boundary conditions:
\begin{align}
H&=-\frac{\hbar^2}{2m}\Delta+\frac{e^2} 2  V
\end{align}
where $V$ is the 2-body Coulomb potential $\sum_{i\neq j}1/|r_i-r_j|$, the electron mass is $m$, and $e$ is its charge.
It is convenient to choose Hartree as the unit of energy, $Ha=\hbar^2/(ma_B^2)$.
We get:
\begin{align}
H&=\frac{a_B^2}2 (-\Delta+ \frac 1 {a_B}V)
\end{align}
Let $\psi_n$ be an orthonormalized set of $N$ vectors of $L^2(\Omega)$. 
They define the $N$-particle Slater determinant $\Psi=\bigwedge_{n} \psi_n$. 
And the energy of $\Psi$ is:
\begin{align}
\label{energy}
{\cal E}=\left<\Psi | H | \Psi\right>&=\frac{a_B^2}{2} \left(-\sum_{n}\left<\psi_n| \Delta | \psi_n\right>+ \frac {1} {a_B}\sum_{n,n'}\left<\psi_n\wedge\psi_{n'} | v | \psi_n\wedge\psi_{n'}\right>\right)
\end{align}
where $v$ is defined as:
\begin{align}
\left<\ffi_1\otimes\ffi_2|v|\psi_1\otimes\psi_2\right>&=
\int dx\,dy \ \overline{\ffi_1}(x) \overline{\ffi_2}(y)\frac{1}{||x-y||}\psi_1(x)\psi_2(y).
\end{align}
In order to avoid problems due to the Coulomb singularity, we  introduce the jellium model and  define the potential acting on the plane waves $\phi_k$ as:
\begin{align}
\left<\phi_k\otimes\phi_{k'}|v|\phi_{k+q}\otimes\phi_{k'-q}\right>&=\frac{\pi}{\Omega}\left(\frac{2}{|q|}\right)^{D-1}
\end{align}
for $q\neq 0$ and $0$ otherwise, so that the total charge of the electrons is compensated by a positive background charge.

The Fermi gas is  defined by $\Phi=\bigwedge_{|k|<k_F} \phi_k$ where $(\alpha_Dk_F)^D=(2\pi)^D N/\Omega$ and $\alpha_D^D$ is the volume of the unit sphere.
\begin{align}
{\cal E}_{FG}=\left<\Phi|H|\Phi\right>&=\frac{a_B^2}{2} \left(\sum_{|k|<k_F}k^2- \frac {2^{D-1}\pi} {a_B\Omega}\sum_{|k|,|k'|<k_F}\vkk\right)
\end{align}
As $\Omega$ goes to $\infty$ with  $\Omega/N$ fixed, the thermodynamic limit for the energy per particle is obtained by the substitution $\sum_k \to \frac{\Omega}{(2\pi)^D} \int dk$:
\begin{align}
\nonumber
\frac{{\cal E}_{FG}}N&=\frac{a_B^2}2 \frac{\Omega}{N(2\pi)^D}\left(\int_{|k|<k_F}dk\ k^2- \frac 1 {a_B2\pi^{D-1}}\int_{|k|,|k'|<k_F}dkdk'\ \vkk\right)\\
&=\frac{a_B^2}2 \frac{\Omega}{N(2\pi)^D}k_F^{D+2}\left(\int_{|k|<1}dk\ k^2- \frac 1 {a_Bk_F2\pi^{D-1}}\int_{|k|,|k'|<1}dkdk'\ \vkk\right)
\end{align}
From the definition of $r_s=(\alpha_D a_B n^{1/D})^{-1}$ and $k_F$, it follows that $k_F\alpha_D^2r_s a_B=2\pi$.
Thus, we have:
\begin{align}
\frac{{\cal E}_{FG}}N=\frac{2\pi^2}{\alpha_D^{D+4} r_s^2}\left(\int_{|k|<1}dk\ k^2- \frac {r_s\alpha_D^2} {4\pi^D}\int_{|k|,|k'|<1}dkdk'\ \vkk\right)
\end{align}
which gives for $D=2$  ($\alpha_2^2=\pi$):
\begin{align}
\label{efg}
\frac{{\cal E}_{FG}}N=\frac{2}{\pi r_s^2}\left(\int_{|k|<1}dk\ k^2- \frac {r_s} {4\pi}\int_{|k|,|k'|<1}dkdk'\ \Vkk\right)
\end{align}
\section{Hartree-Fock upper bounds for the polarized 2D electron gas}
\label{bounds}
In this section we estimate the energy for a class of states inspired by our numerical results.

Let us consider a state $\Psi=\bigwedge_{|k|<k_F} \psi_k$
where:  
\begin{align}
\label{psidef}
\psi_k=a_k\phi_k+b_k\phi_{k+Q_k}
\end{align}
with $Q_k$ in $\{-2k_F(\cos p\pi/3,\sin p\pi/3)\}_{p=0\ldots 5}$.
For $k=|k|(\cos \theta, \sin \theta)$ we choose $Q_k$ such that $|k+Q_k|$ is minimal; that is, we choose $p$ as the integer part of $(3\theta/\pi+1/2)$ and we must assume $b_k$ is zero if $k$ is zero or $\theta=\pi/6+n\pi/3$.

\begin{figure}
\begin{center}
\includegraphics[width=0.4\textwidth]{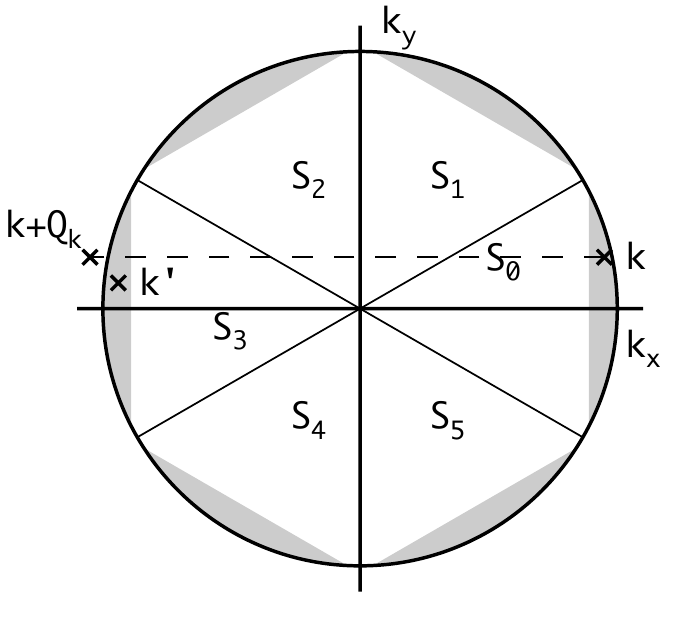}
\caption{The circle is the Fermi surface. The shaded surfaces are the regions where $b(k)$ is nonzero. The new state $\psi_k$
mixing $\phi_k$ and $\phi_{k+Q_k}$ is now resonant with $\psi_{k'}$.
}
\label{resonant}
\end{center}
\end{figure}

Furthermore, we assume that $a_k$ and $b_k$ are real positive number and invariant thru the rotation of $2n\pi/6$ and the symmetry $\theta\rightarrow -\theta $ (i.e., the dihedral group $D_6$). 
The $ \psi_k$'s are normalized, so that $a_k^2+b_k^2=1$ and  $b_k=0$ if $|k\cdot Q_k|<2k_F^2(1-\epsilon)$ (i.e., $b_k$ is not zero only in the vicinity of $\{k_F(\cos p\pi/3,\sin p\pi/3)\}_{p=0\ldots 5}$), see Fig.\ref{resonant}. 

Thus, from Eqs \ref{energy} and \ref{psidef}, the limit energy per particle is given by: 
\begin{align}
\frac{{\cal E}}N=\frac{2}{\pi r_s^2}\left(\int_{|k|<1}dk\ \left<\psi_k|-\Delta | \psi_k\right>+ \frac {r_s} {4\pi}\int_{|k|,|k'|<1}dkdk'\ \frac{\Omega}{2\pi}\left<\psi_k\wedge\psi_{k'}|v| \psi_k\wedge\psi_{k'}\right>\right)
\end{align}
where, as in (\ref{efg}),  the $k$'s have been renormalized by $k_F$ and thus $|Q_k|=2$.

We define $\Delta E$ by:
\begin{align}
\label{deltaEdef0}
\frac{{\cal E}-{\cal E}_{FG}}N=\frac{2}{\pi r_s^2}\Delta E
\end{align}
Then
\begin{align}
\label{defde}
\Delta E=\int_{|k|<1}dk\,\left[\left<\psi_k|-\Delta| \psi_k\right>-k^2\right]+ \frac {r_s} {4\pi}\Delta E_V
\end{align}
where
\begin{align}
\label{dev}
\Delta E_V=\int_{|k|,|k'|<1}dkdk'\, \left(\frac{\Omega}{2\pi}\left<\psi_k\wedge\psi_{k'}|v | \psi_k\wedge\psi_{k'}\right>+\Vkk\right)
\end{align}
\subsection{Potential energy contribution: $\Delta E_V$}
Setting $v_q=1/|q|$:
\begin{align}
\nonumber
\frac{\Omega}{2\pi}\left<\psi_k\wedge\psi_{k'}|v| \psi_k\wedge\psi_{k'}\right>+v_{k-k'}&=(v_{k-k'}-v_{k-k'-Q_{k'}})b_{k'}^2a_k^2\\
\nonumber
&+(v_{k-k'}-v_{k+Q_{k}-k'})b_{k}^2a_{k'}^2\\
\nonumber
&+(v_{k-k'}-v_{k+Q_{k}-Q_{k'}-k'})b_{k}^2b_{k'}^2\\
\nonumber
&+2v_{Q_k}a_k b_k  a_{k'} b_{k'}(\delta_{Q_k+Q_{k'}}+\delta_{Q_k-Q_{k'}})\\
\nonumber
&-2v_{k-k'}a_k b_k  a_{k'} b_{k'}\delta_{Q_k-Q_{k'}}\\
\label{ve0plusv}
&-(v_{k+Q_k-k'}+v_{k-k'-Q_{k'}})a_{k} a_{k'}  b_k b_{k'}\delta_{Q_k+Q_{k'}}
\end{align}
Eq.(\ref{dev}), may be divided into 4 parts:
\begin{itemize}
\item \underline{\{$b_{k'}=0$, $b_k= 0$\}:} the contribution is zero. 
\item \underline{\{$b_{k'}=0$, $b_k\neq 0$\}, \{$b_{k'}\neq 0$, $b_k= 0$\}:} both cases are equivalent. 

For \{$b_{k'}=0$, $b_k\neq 0$\}, the integrant of Eq. \ref{dev} is:
\begin{align}
\frac{\Omega}{2\pi}(\psi_k\wedge\phi_{k'},v\ \psi_k\wedge\phi_{k'})+\Vkk=(v_{k-k'}-v_{k+Q_{k}-k'})b_k^2
\end{align}
Let  $S_0$ be the sector of unit disk between $-\pi/6$ and $\pi/6$ (see Fig. \ref{resonant}); then in this sector $Q_{k}=(-2,0)$ and by symmetry:
\begin{align}
\int_{b_{k'}=0}dkdk'\ (v_{k-k'}-v_{k+Q_{k}-k'})b_k^2
&=6\int_{k\in S_0, b_{k'}=0}dkdk'\ (v_{k-k'}-v_{k+Q_{k}-k'})b_k^2\\
&=6\int_{k\in S_0, b_{k'}=0}dkdk'\ (v_{k-k'}-v_{\tilde k-k'})b_k^2\\
&\leq C\epsilon^3+6\int_{k\in S_0, |k'_x|<1-\epsilon}dkdk'\ (v_{k-k'}-v_{\tilde k-k'})b_k^2
\end{align}
where $\tilde k=(2-k_x,k_y)$.
In $S_0$, $k=(k_x,k_y)$ where $k_x$ is close to 1 and setting $k_x=1-x$, we assume from now that $b_k=b(x/\epsilon)$.

In Appendix A, we prove that:
\begin{align}
\label{bk0}
\int_{k\in S_0,|k'_x|<1-\epsilon}dkdk'\ (v_{k-k'}-v_{k+Q_{k}-k'})b_k^2\leq 8\epsilon^2\sqrt{2\epsilon}\left[\ln\epsilon^{-1}+O(1)\right]\int_0^1 dx b^2(x)x\sqrt x
\end{align}
\item \underline{\{$b_{k'}\neq0$, $b_k\neq 0$\}:}

By symmetry we can assume that $k$ belongs to $S_0$. If $k'\not \in S_0\cup S_3$ all the $v$ appearing in (\ref{ve0plusv}) are uniformly  bounded.
And since the k-volume for each sector goes like $\epsilon\sqrt\epsilon$, the contribution of these terms is bounded by $C\epsilon^3$.
In the same way $v_{k-k'}$ is bounded when $k'\in S_3$ and $v_{k+Q_k-k'}$ is bounded when $k'\in S_0$.
Thus setting:
\begin{align}
f&:=a_k^2b_{k'}^2+b_k^2a_{k'}^2-2a_{k} a_{k'}  b_k b_{k'}=(a_kb_{k'}-b_k a_{k'})^2\\
g&:=a_k^2b_{k'}^2+b_k^2a_{k'}^2+2a_ka_{k'}b_{k} b_{k'}=(a_kb_{k'}+b_k a_{k'})^2
\end{align}
one can check that:
\begin{align}
\label{dev2}
\int_{b_k,b_{k'}\neq 0}\frac{\Omega}{2\pi}\left<\psi_k\wedge\psi_{k'}|v| \psi_k\wedge\psi_{k'}\right>+v_{k-k'}\leq C\epsilon^3+6\int_{k,k'\in S_0}dkdk'\ \left(v_{k-k'}f-v_{k+Q_k+k'}g\right)
\end{align}
In Appendix B we prove that
\begin{align}
\label{bk1}
\int_{k,k'\in S_0}dkdk'\ \left(v_{k-k'}f-v_{k+Q_k+k'}g\right)\leq 4\epsilon^2\sqrt{2\epsilon}\left[\ln\epsilon^{-1}+O(1)\right]\int_0^1 dx\sqrt x\int_x^1 dx' \left(f( \epsilon x, \epsilon x')-g( \epsilon x, \epsilon x')\right)
\end{align}
\end{itemize}
Thus, summing the four contribution gives:
\begin{align}
\label{dev3}
\Delta E_V&\leq C\epsilon^3+6\epsilon^2\sqrt{2\epsilon}\left[\ln\epsilon^{-1}+O(1)\right]\int_0^1 dx\sqrt x\left(16b^2(x)x+4\int_x^1 dx'\ \left(f(x, x')-g( x, x')\right)\right)
\end{align}

\subsection{Kinetic energy contribution:}
The variation of the kinetic energy is given by:
\begin{align}
\int_{|k|<1}dk\ \left[\left<\psi_k | -\Delta | \psi_k\right>-k^2\right]&=6\int_{k\in S_0} dk (\left<\psi_k|-\Delta| \psi_k\right>-k^2)\\
&=6\int_0^\epsilon dx\ 2y_m\ 4xb^2(x/\epsilon)\\
\label{kin}
&\leq 6\times 8\epsilon^2\sqrt{2\epsilon}\int_0^1 dx\ \sqrt x\ xb^2(x)
\end{align}

\subsection{Total energy:}
Inserting Eqs (\ref{dev3},\ref{kin}) in Eq. (\ref{defde}), the variation of the total energy from the Fermi gas energy becomes:
\begin{align}
\nonumber
\Delta E&\leq 6\epsilon^2\sqrt{2\epsilon}\int_0^1 dx\sqrt{x}\left(8xb^2( x)+\frac {r_s} {4\pi}\left[\ln\epsilon^{-1}+O(1)\right]\left(16b^2( x)x+4\int_x^1 dx'\left(f(x, x')-g(x, x')\right)\right)\right)\\
\label{def}
&= 6\times 8\epsilon^2\sqrt{2\epsilon}\int_0^1 dx\sqrt{x}\left(xb^2( x)+\frac {r_s} {2\pi}\left[\ln\epsilon^{-1}+O(1)\right]\left(b^2( x)x-a(x)b(x)\int_x^1 dx'a(x')b(x'))\right)\right)
\end{align}
Let us set 
\begin{align}
\label{defdelta}
\delta&=\epsilon^2\sqrt{\epsilon}\\
\label{i1i2-1}
I_1&=\int_0^1 dx\sqrt xxb^2( x)\\
\label{i1i2bis-1}
I_2&=\frac{1}{5\pi}\int_0^1 dx\sqrt x\left(-b^2( x)x+a(x)b(x)\int_x^1 dx'a(x')b(x')\right)
\end{align}
Then
\begin{align}
\label{deltaEdef}
\Delta E&\leq  6\times 8\sqrt{2}\delta\left[I_1-r_sI_2(\ln \delta ^{-1}+O(1))\right]
\end{align}
If $I_2>0$, as $r_s$ goes to $0$, $\Delta E$ is minimal in Eq. (\ref{deltaEdef}) for  $\delta$ defined by:
\begin{align}
\label{deltamin}
\delta_{\min}=\frac 1 {e}\exp\left(-\frac {I_1}{I_2r_s}\right)
\end{align}
and finally inserting $\delta_{\min}$ in Eq. (\ref{deltaEdef}) gives:
\begin{align}
\label{DEF1}
\Delta E\lesssim-\frac{6\times 8\sqrt{2}}{e}\exp\left(-\frac {I_1}{I_2r_s}\right)r_sI_2
\end{align}
We now have to find a solution $b(x)$ such that $I_2$ is positive. Choosing $b(x)=b_0$ or $b(x)=b_0 ( 1 - x ) $ leads to negative $I2$. 
In the Appendix C, as $r_s$ goes to 0 we find a family of $b$ leading to :
\begin{align}
\label{deb}
\Delta E&\lesssim -r_s\exp\left(-\frac {5\pi}{3r_s}+\frac {O(1)} {\sqrt {r_s}}\right)
\end{align}
Though such a bound is correct in the thermodynamic limit for $r_s \to 0$, this behavior in not so relevant for finite systems. 
Our numerical calculations of section \ref{numerics} consider about $10^3$ electrons where the uniform Fermi gas remains  the ground state for $r_s \lesssim 1$.
Thus the asymptotic bound (\ref{deb}) is not very helpful in comparing with our numerical results obtained for $r_s \approx 1$.

Nevertheless, for finite $r_s$, on can choose a suitable function $b$ and evaluate numerically $I_1$ and $I_2$.
For instance, with $b=b_\eta$ as in (\ref{beta}) of Appendix C and $\eta=0.001$ we get 
\begin{align}
\frac{{\cal E}-{\cal E}_{FG}}N&\lesssim-2.\times 10^{-4}r_s^{-1}\exp\left(-\frac { 18.5}{r_s}\right)
\end{align}

We can also understand why the metallic phase does not neccessarily appear in small sized systems.
For finite systems of $N$ electrons, one must have at least one plane wave in the shaded region of figure \ref{resonant}: $|k \cdot Q_k|>2k_F^2(1-\epsilon)$.
This gives the condition
$N\epsilon\sqrt\epsilon>1$ and using Eqs (\ref{defdelta},\ref{deltamin}), this leads $N>\exp(3I_1/5I_2 r_s)$.
Analogous to Eq. (\ref{deb}), we obtain the following  lower bound
\begin{align}
N>\exp\left(\frac{3 \pi}{ r_s}\right)
\end{align}
i.e. $N>500$ for $r_s=1.8$. This bound is compatible with our numerical simulations where the metallic phase disappears at $r_s=1$ for $N=500$, and
may explain why the metallic phase has not been observed in previous numerical calculations.

\section{Conclusion}
\label{conclusion}
Using a descent algorithm, we have computed the ground state of up to $N=500$ electrons.
For $1\lesssim r_s\lesssim 3$, our solutions have lower energies than the FG or WC.
These solutions correspond to denser lattices that the WC solutions, that is with less than one electron per site as in a metallic material.

We have proven, in the thermodynamic limit, that for sufficiently small $r_s$, these metallic states have always a smaller energy than the Fermi gas. 
To our knowledge, it is the first time that rigorous  upper bounds for the ground state energy of the polarized electron gas are obtained, demonstrating that the FG is not the ground state even at small $r_s$.

\begin{figure}
\begin{center}
\includegraphics[width=0.45\textwidth]{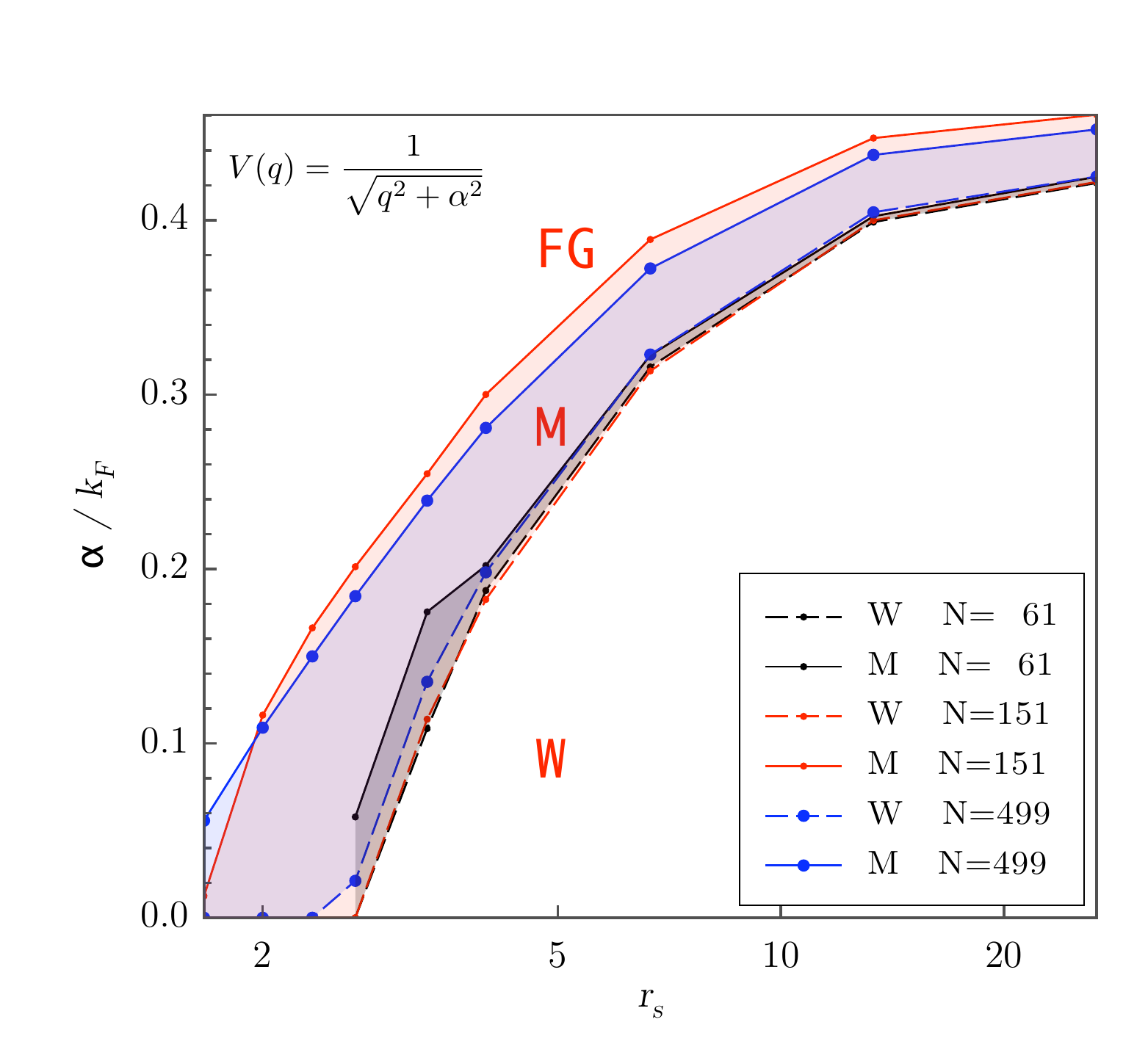}
\caption{Effect of the screening on the ground state phase diagram for $N=61$, $151$, and $499$. The different
phases are labelled
``FG" ( Fermi-gas ground state), ``M" (metallic ground state), and  ``W"  ( Wigner crystal).
For all values of the screening parameter $\alpha$ of the screened Coulomb
potential $V(q)$, we have found a metallic phase of energy lower than the Fermi gas 
and the Wigner crystal energy.}
\label{screenedpot}
\end{center}
\end{figure}

Our proof relies on the behavior at infinity of the Coulomb potential, so it may be interesting to check the existence of these states
in the case of a screened Coulomb potential. A rigorous extension of our proof is not straightforward. However, 
as shown in Fig.~\ref{screenedpot},
our numerical calculations indicate that
 the metallic phase persists in the presence of screening, at least for various system sizes and screening 
 parameters studied. Thus, such metallic states 
should be considered as relevant candidates for further studies beyond the Hartree-Fock approximation,
since, qualitatively, correlation effects  amounts to an effective screening
 of the electron interaction in the high density limit.

\section{Appendix A}
We have to estimate 
\begin{align}
I(f)=\int_{\begin{subarray}{1} \|k\|,\|k'\|\leq 1\\1-k_x<\epsilon,|k'_x|<1-\epsilon\end{subarray}}dkdk'\ (v_{k-k'}-v_{\tilde k-k'})f(1-k_x)
\end{align}
where $\tilde k=(2-k_x,k_y)$ and $f$ is a positive function.
\begin{align}
\int dk'\ (v_{k-k'}-v_{\tilde k-k'})=\int dk_x'\ \asinh\frac{y'_m-k_y}{k_x-k'_x}+\asinh\frac{y'_m+k_y}{k_x-k'_x}- \asinh\frac{y'_m-k_y}{2-k_x-k'_x}-\asinh\frac{y'_m+k_y}{2-k_x-k'_x}
\end{align}
where $y'_m=\sqrt{1-{k'_x}^2}$. And since $\asinh x -\asinh y\leq\ln x/y$ for $x>y>0$:
\begin{align}
\int dk'\ (v_{k-k'}-v_{\tilde k-k'})\leq\int_{-1+\epsilon}^{1-\epsilon} dk_x'\  2\ln\frac{2-k_x-k'_x}{k_x-k'_x}
\end{align}
We set $k_x=1-x$ and $y_m=\sqrt{2x-x^2}$,
\begin{align}
I(f)&\leq\int_0^\epsilon dx f( x)\ 2y_m
\int_{-1+\epsilon}^{1-\epsilon} dk_x'\  2\ln\frac{1+x-k'_x}{1-x-k'_x}\\
&=4\int_0^\epsilon dx f( x)\ y_m\int_{\epsilon}^{2-\epsilon} du\  \ln\frac{u+x}{u-x}\\
&\leq 4\int_0^\epsilon dx f( x)\ y_m\int_{\epsilon}^{2-\epsilon} du\ \frac{2x}{u-x}\\
&=8\epsilon^2\sqrt{2\epsilon}\left[\ln\epsilon^{-1}+O(1)\right]\int_0^1 dx f(\epsilon x)x\sqrt x
\end{align}

\section{Appendix B}
We have to estimate
\begin{align}
\label{dvdef}
I(f,g)=\int_{\begin{subarray}{1} \|k\|,\|k'\|\leq 1\\1-k_x,1-k'_x<\epsilon\end{subarray}}dkdk'\ \left(v_{k-k'}f-v_{\tilde k+k'}g\right)
\end{align}
where $\tilde k=(k_x-2,k_y)$ and $f$ and $g$ are positive functions of $1-k_x$ and $1-k_{x'}$.
Setting $k_x=1-x$, $k'_x=1-x'$, $k_y=y$,  $k'_y=y'$, and $r_\pm=\sqrt{(x\pm x')^2+(y-y')^2}$, Eq.\ref{dvdef} can be rewritten:
\begin{align}
\nonumber
\nonumber
I(f,g)&=\int_0^\epsilon dx\int_0^\epsilon dx' \int dydy'\left(\frac 1{r_-}f-\frac 1{r_+}g\right)\\
\label{dv1}
&=2\int_0^\epsilon dx\int_x^\epsilon dx' \int dydy'\left(\frac 1{r_-}f-\frac 1{r_+}g\right)
\end{align}
where $y$ and $y'$ must satisfy $(1-x)^2+y^2\leq 1$ and $(1-x')^2+{y'}^{2}\leq 1$.\\
Since $\asinh x\leq \ln 2(x+1)$, the first term in \ref{dv1} is bounded by:
\begin{align}
2\int_0^\epsilon dx\int_x^\epsilon dx' \int dydy'\ \frac 1{r_-}f&=2\int_0^\epsilon dx\int_x^\epsilon dx' f\int_{-y_m}^{y_m} dy\ (\asinh \frac{y'_m+y}{x'- x}+\asinh \frac{y'_m-y}{x'- x})\\
&\leq2\int_0^\epsilon dx\int_x^\epsilon dx' f\int_{-y_m}^{y_m} dy\ 2\asinh \frac{2y'_m}{x'- x}\\
&\leq 4\int_0^\epsilon dx\int_x^\epsilon dx' f2y_m\ln (2+\frac{4y'_m}{x'- x})\\
&\leq 4\epsilon^2\sqrt{2\epsilon}\left[\ln(\epsilon^{-1})+O(1)\right] \int_0^1 dx\int_x^1 dx\,f( \epsilon x, \epsilon x')\sqrt x
\end{align}
On the other hand, using $\asinh x\geq \ln 2x$, the last term of (\ref{dv1}) is:
\begin{align}
2\int_0^\epsilon dx\int_x^\epsilon dx' g\int dydy'\frac 1{r_+}
&=2\int_0^\epsilon dx\int_x^\epsilon dx' g\int_{-y_m}^{y_m} dy\asinh \frac{y'_m-y}{x+x'}+\asinh \frac{y'_m+y}{x+x'}\\
&\geq 2\int_0^\epsilon dx\int_x^\epsilon dx' g\int_{-y_m}^{y_m} dy\ln 4\frac{{y'_m}^2-y^2}{(x+x')^2}\\
&\geq 4\int_0^\epsilon dx\int_x^\epsilon dx' g y_m\left[\ln\epsilon^{-1}+O(1)\right]\\
&\geq 4\epsilon^2\sqrt{2\epsilon}\left[\ln\epsilon^{-1}+O(1)\right]\int_0^1 dx\int_x^1 dx' g(\epsilon x, \epsilon x')\sqrt x
\end{align}
And we have:
\begin{align}
I(f,g)&\leq 4\epsilon^2\sqrt{2\epsilon}\left[\ln\epsilon^{-1}+O(1)\right]\int_0^1 dx\sqrt x\int_x^1 dx' \left(f( \epsilon x, \epsilon x')-g( \epsilon x, \epsilon x')\right)
\end{align}

\section{Appendix C}
Here we provide exact bounds on $I_1$ and $I_2$ given by (\ref{i1i2-1}, \ref{i1i2bis-1}).

In order to estimate $I_2$ we introduce the linear operator $A$ :
\begin{align}
Af(x)&=\frac 1 {2x}\int_x^1f(y)dy+\frac 1 {2x\sqrt x }\int_0^xf(y)\sqrt {y} dy
\end{align}
defined on the Hilbert space of the functions on $[0,1]$ with the scalar product:
\begin{align}
\left<f|g\right>&=\int_0^1 x\sqrt x\,\overline{f(x)}g(x) dx
\end{align}
Then $A$ is a bounded symmetric operator and:
\begin{align}
\label{i2i1}
I_2/I_1=\frac{1}{5\pi}\left(\frac{\left<ab|Aab\right>}{\| b\|^2}-1\right)
\end{align}
The unitary operator $f(x)\rightarrow g(y)=f(e^{-y})e^{-5/4y}$ from $L^2([0,1],x\sqrt x dx)$ onto $L^2([0,+\infty],dx)$ maps the operator $A$ onto the operator $\tilde A$:
\begin{align}
\tilde Ag(x)&=\frac {e^{-x/4}} 2 \int _0^xe^{y/4}g(y)dy +\frac{e^{x/4}} 2 \int _x^{+\infty}e^{-y/4}g(y)dy
\end{align}
Then 
\begin{align}
\tilde Ae^{ikx}&=\frac{1}{4(1/16+k^2)}e^{ikx}-\frac 1 {1/2+i2k}e^{-x/4}
\end{align}
Thus setting:
\begin{align}
g_k(x)=\frac 1 {|1+i4k|}\left[(1+i4k)e^{ikx}-(1-i4k)e^{-ikx}\right]
\end{align}
$\{g_k\}_{k>0}$ is  a full set of pseudo-eigenvectors satisfying: $$\tilde Ag_k=\frac{1}{4(1/16+k^2)}g_k$$
Thus the spectrum of $\tilde A$ is $(0,4)$ and the spectral measure is purely absolutely continuous;  the largest spectral value is $4$ with a pseudo-eigenvector $g_4(x)=x+4$ corresponding to $f_4(x)=x^{-5/4}(4-\ln x)$.
But $\|f_4\|$ is infinite and $f_4$ diverges at $0$. 
The next step is to choose a family of functions $b_\eta$ such that $a_\eta=\sqrt{1-b_\eta}$ is defined and $\left<a_\eta b_\eta|Aa_\eta b_\eta\right>/\| b_\eta\|^2$ is close to 4.

Thus setting $f_{\eta}(x)=\min(f_4(x),f_4(\eta))$ for $0<\eta\ll 1$, we have:
\begin{align}
\|f_{\eta}\|^2&=-\frac 1 3\left[ \ln^3 \eta-\frac{66} 5 \ln^2 \eta +O(\ln \eta)\right]\\
\left<f_{\eta}|Af_{\eta}\right>&=-\frac 4 3\left[\ln^3 \eta-\frac {41} {5} \ln^2 \eta +O(\ln \eta)\right]
\end{align}
Then:
\begin{align}
\label{Afeta}
\frac{\left<f_{\eta}|Af_{\eta}\right>}{\|f_{\eta}\|^2}&=4-\frac {20}{| \ln \eta|}+O(\ln^{-2} \eta)
\end{align}
Thus $f_\eta$ is a good candidate for the linear part of the problem. Now, by the simple scaling:
\begin{align}
\label{beta}
b_{\eta}(x)=\frac{f_{\eta}(x)}{\sqrt 2f_{\eta}(\eta)}
\end{align}
we get the nonlinear candidate satisfying $b_{\eta}(x)\leq1/\sqrt 2$, $a_{\eta}=\sqrt{1-b_{\eta}^2}$ is well defined, $a_{\eta}(x)\geq1/\sqrt 2$ and $b_{\eta}$ satisfies (\ref{Afeta}).

We must now estimate the simultaneous convergence of $I2/I1$ (\ref{i2i1}) and $I_2$ as $\eta$ decreases.
\begin{align}
\nonumber
\left<b_{\eta}|Ab_{\eta}\right> -\left<b_{\eta}a_{\eta}|Ab_{\eta}a_{\eta}\right>&=-\left<b_{\eta}-b_{\eta}a_{\eta}|A|b_{\eta}-b_{\eta}a_{\eta}\right>
+2\left<b_{\eta}-b_{\eta}a_{\eta}|A|b_{\eta}\right>\\
\nonumber
&\leq 2\left<b_{\eta}-b_{\eta}a_{\eta}|A|b_{\eta}\right>\\
\nonumber
&= 8\left<b_{\eta}-b_{\eta}a_{\eta}|b_{\eta}\right>+2\left<b_{\eta}-b_{\eta}a_{\eta}|(A-4)b_{\eta}\right>\\
\nonumber
&\leq 8\left<b_{\eta}-b_{\eta}a_{\eta}|b_{\eta}\right>+2\|b_{\eta}-b_{\eta}a_{\eta}\|\| (A-4)b_{\eta}\|\\
\nonumber
&\leq 8\left<b_{\eta}-b_{\eta}a_{\eta}|b_{\eta}\right>+8\|b_{\eta}-b_{\eta}a_{\eta}\|\sqrt{\left< b_{\eta}|(A-4)b_{\eta}\right>}
\end{align}
where:
\begin{align*}
\left<b_{\eta}-b_{\eta}a_{\eta}|b_{\eta}\right>&=\int_0^1b_{\eta}(x)^2\left[1-a_{\eta}(x)\right]x\sqrt xdx\\
&\leq \int_0^1b_{\eta}(x)^2\left[1-a_{\eta}(x)\right]^2x\sqrt{x}dx \sup \frac 1 {1- a_{\eta}}\\
&\leq \|b_{\eta}-b_{\eta}a_{\eta}\|^2 \frac {\sqrt 2}{\sqrt 2-1}\\
\end{align*}
and since $b_{\eta}-b_{\eta}a_{\eta}>0$
\begin{align*}
\|b_{\eta}-b_{\eta}a_{\eta}\|^2&=\|b_{\eta}\|^2-\|b_{\eta}a_{\eta}\|^2-2\left<b_{\eta}-b_{\eta}a_{\eta}|b_{\eta}a_{\eta}\right>\\
&\leq \|b_{\eta}\|^2-\|b_{\eta}a_{\eta}\|^2\\
&= \|b_{\eta}^2\|^2
\end{align*}
By a direct computation: $$\|b_{\eta}^2\|^2\leq \|b_{\eta}\|^2 \frac{6}{5|\ln \eta|}$$ for $\eta$ small enough, and thus:
\begin{align}
\label{dbAb}
\frac{\left<b_{\eta}|Ab_{\eta}\right>}{\|b_{\eta}\|^2} -\frac{\left<b_{\eta}a_{\eta}|Ab_{\eta}a_{\eta}\right>}{\|b_{\eta}\|^2} \leq  \frac{8}{|\ln \eta|}\left(\frac {\sqrt 2}{\sqrt 2-1}\frac 6 5+2\sqrt 6\right)
\end{align}
And finally, from (\ref{Afeta}) for $b_{\eta}$ and (\ref{dbAb}) and $I_1= \|b_{\eta}\|^2$,   (\ref{i2i1}) gives:
\begin{align}
I_2/I_1&\geq\frac{1}{5\pi}\left(3-\frac{C}{|\ln \eta|}\right)+O(\ln^{-2} \eta)\\
I2&\leq \frac 2 {15\pi} \eta^{5/2}|\ln^3 \eta|\left[1+O\left(\frac{1}{|\ln \eta|}\right)\right]
\end{align}
where
\begin{align}
C=20+8\left(\frac {\sqrt 2}{\sqrt 2-1}\frac 6 5+2\sqrt 6\right)\approx 92
\end{align}
Choosing $\eta$ sufficiently small, this proves that $\Delta E$ is strictly negative for any $r_s>0$; furthermore choosing $\eta$ to minimize $\Delta E$ (Eq. \ref{DEF1}),  i.e., $9r_s|\ln \eta|^2=2C\pi$, we obtain as $r_s$ goes to 0:
\begin{align}
\Delta E&\lesssim -r_s\exp\left(-\frac {5\pi}{3r_s}+\frac {O(1)} {\sqrt r_s}\right)
\end{align}

Acknowledgment: We thank J. Trail and R. Needs for providing us numerical data of Ref.~\cite{Needs} shown in Fig.\ref{FIG-Comparison}. We thank D. Ceperley for discussions.

\end{document}